\documentclass{jltp}
\usepackage{graphicx} 

  \def\qra {\rangle}

\def\Vph#1{\widetilde V_{\mbox{\footnotesize p-h}}(#1)\,}
\def\Vphe{\widetilde V_{\mbox{\footnotesize p-h}}\,}

\def\he#1{$^#1$He}

\title{Pair Excitations and Vertex Corrections in Fermi Fluids}

\author{Helga M. B\"ohm, Eckhard Krotscheck, and Martin Panholzer}

\address{Inst.\ f.\ Theoretische Physik,
         Johannes Kepler Universit\"at, A-4040 Linz, Austria}
\runninghead{H.~M.~B\"ohm,~E.~Krotscheck, and M.~Panholzer}
            {Pair Excitations in Fermi Fluids}

\begin{document}

\maketitle

\begin{abstract}
Based on an equations--of--motion approach for time--dependent {\it pair\/}
correlations in strongly interacting Fermi liquids, we have developed a
theory for describing the excitation spectrum of these systems. Compared to
the known ``correlated'' random--phase approximation (CRPA), our
approach has the following properties:
i) The CRPA is reproduced when pair fluctuations are neglected.
ii) The first two energy--weighted sumrules are fulfilled implying a correct
static structure.
iii) No ad--hoc assumptions for the effective mass are needed
to reproduce the experimental dispersion of the roton in $^3$He.
iv) The density response function displays a novel form, arising from
vertex corrections in the proper polarisation.
Our theory is presented here with special emphasis on this latter point.
We have also extended the approach to the single particle self-energy
and included pair fluctuations in the same way. The theory provides
a diagrammatic superset of the familiar GW approximation. It
aims at a consistent calculation of single particle
excitations with an accuracy that has previously only been achieved
for impurities in Bose liquids.
\end{abstract}



\section{MOTIVATION}
\label{sec:motivation}

Whereas the ground state properties of strongly interacting fermionic
fluids are nowadays well understood, the features of higher-lying
excitations such as the roton in He and the plasmon in electron liquids
still require further clarification.  Without a proper consideration
of pair excitations, both the dispersion and the damping of these
collective modes are described incorrectly.  In the bosonic case, in
particular for the roton in $^4$He, the introduction of ``backflow''
correlations\cite{FeynmanBackflow} (equivalent to a special case of
pair excitations) has led to a
theory\cite{JacksonSumrules,VesaMikkou2} yielding remarkable agreement
with experiment.  The correction of the roton position is due to pair
fluctuations which have wavelengths comparable to the interparticle
distance.

In Fermi liquids the density-density response function is conventionally
taken to be of the CRPA form
\begin{equation}
\chi^{{\rm CRPA}}(q,\omega)
= {\chi_0(q,\omega)/\bigl( 1-\Vph{q}\chi_0(q,\omega)\bigr)} \;,
\label{eq:chirpa}
\end{equation}
where $\chi_0$ is the Lindhard function and $\Vphe$ a suitable
effective interaction (also termed ,,pseudopotential\cite{IP84}\,'' and 
,,local field\cite{SiT81} corrected potential''). 
Fixing $\Vphe$ through the $\omega^0$ and $\omega^1$ sum rules
leads to a collective mode that is energetically much higher than
found in experiments. One can cure this problem in an {\it ad-hoc\/}
manner by introducing an average effective mass\cite{GFvDG00} $m^*$ in
$\chi_0$. This is unsatisfactory from the point of view of developing a
manifestly microscopic description of the low-temperature properties
of \he3, and it also introduces inconsistencies since it violates
the above sum rules. In addition, one would expect that the physical mechanisms
that lead to the lowering of the phonon-roton spectrum in \he4
from the Feynman dispersion law are also at work in \he3.
As in \he4 the phonon-roton spectrum has been understood quantitatively
without the violation of the two above sumrules, our goal is here
to develop a theory for \he3 that has the same level of accuracy and
consistenty as the \he4 theory.

The need for such a theory is made even clearer in
quasi--2--dimensional $^3$He, where the collective mode is found {\it
inside}\/ the particle-hole continuum\cite{GodfrinMeschke}.  An
effective $m^*$ in an RPA formula (\ref{eq:chirpa}) may lower (both!)
the continuum and the collective mode, but only dynamic correlations
could reproduce the broadening of the phonon.

\section{FORMALISM}
\label{sec:form}
\def\smf#1 {\mbox{\small$#1$}} 
\def\symph {\left({\textstyle {{p \,\leftrightarrow\, p' }\atop
                                h \,\leftrightarrow\, h' } }\right)}

\begin{figure}
  \vskip0.1cm
  \includegraphics[width=.98\textwidth]{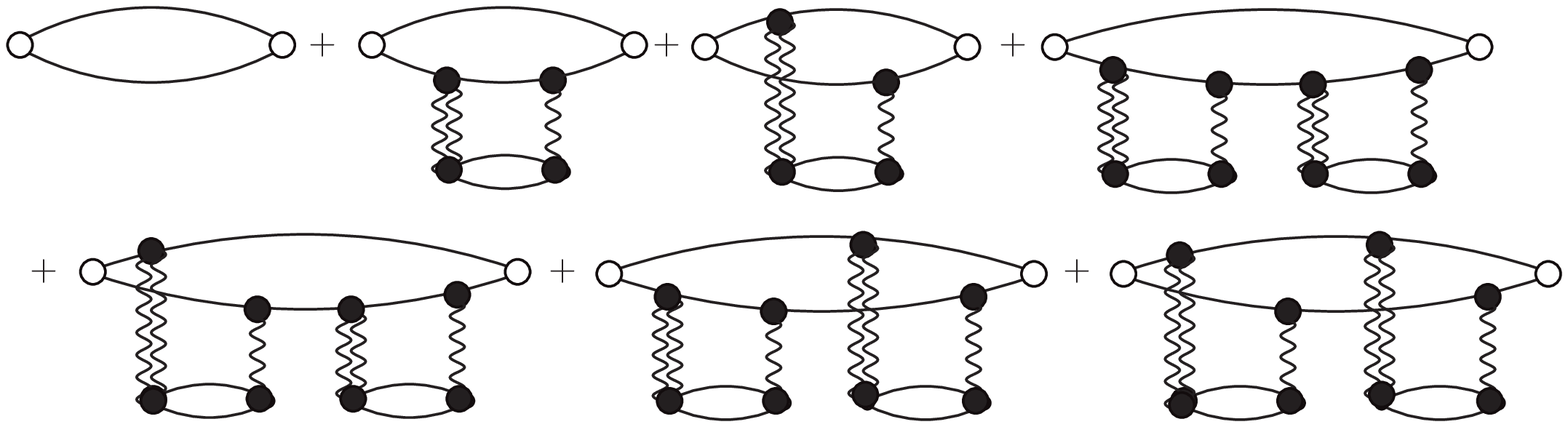}
  \\ \hspace*{4.4cm}\small{$+$} $\ldots$\\ 
  \caption{%
     Typical proper polarization Feynman diagrams taken into account in our 
     approach.
     (The single and double wavy lines denote the bare and screened 
     interactions, respectively. The lines with arrows denote free single 
     particle
     propagators).}
  \label{fig:FD_Pol}
\end{figure}

We describe the excited state wave function of strongly interacting fermions
by a generalized time-dependent Hartree-Fock form
\begin{equation}
 \big|\Psi(t) \qra \>=\> \textstyle\frac{1}{{\cal N}^{1/2}}\,\displaystyle
         e^{-i H_{00}\,t /\hbar}\>
         F\,e^{\,U(t)}\,\big|\Phi_0 \qra \;,
\end{equation}
where  ${\cal N}$ denotes the normalization integral, $\big|\Phi_0\qra$ the 
Slater determinant and $H_{00}$ the energy of the correlated\cite{FeenbergBook}
ground state $F\,\big|\Phi_0\qra$. 
The excitation operator $U$ includes both single-pair (``$c$'') and {\it 
two-pair\/} (``$d$'') amplitudes:
\begin{equation}
 U(t) \>=\> \sum_{ph} c_{ph}(t)\, a^\dagger_p a^{\phantom\dagger}_h +
 \textstyle{1\over 2}\displaystyle\,\sum_{phh'p'}  d_{pp'hh'} (t)\, a^\dagger_p
 a^\dagger_{p'} a^{\phantom\dagger}_{h'} a^{\phantom\dagger}_{h} \;,
\label{eq:FluctU}
\end{equation}
The amplitudes $c_{ph}(t)$ and $d_{pp'hh'}(t)$ are determined by
minimizing the action integral corresponding to the time--dependent
Schr\"odinger equation.  Treating the external field $h^{\rm ext}$ and 
consequently $c_{ph}$ and $d_{pp'hh'}$ as first order perturbations, the 
resulting equations of motion (EOMs) read
\begin{eqnarray}
  \left[i\hbar{\partial\over\partial t} - e_{ph}\right]
  c_{ph}(t) &=& h^{\rm ext}_{ph}(t) 
  + 
    \langle \smf{ph'} |\Vphe| \smf{hp'} \rangle_{\!_a\,} c_{p'h'} +
  \\ \nonumber  
    \langle \smf{pp'} |\Vphe| \smf{hh'} \rangle_{\!_a\,} c_{p'h'}^*
  &+& 
    \langle \smf{ph'}   |W|\smf{p''p'} \rangle_{\!_a\,} d_{p''p',hh'}
  - 
    \langle \smf{h''h'} |W|\smf{hp'}   \rangle_{\!_a\,} d_{pp',h''h'} \;,
\label{eq:1p1hweakb}
\end{eqnarray}
with $e_{ph}\equiv e_p\!-\!e_h$ denoting the particle-hole  excitation energies 
and $W$ the effective interaction in the channels other than particle--hole 
($\widetilde V_{_{\rm p-h}}$), particle--particle ($\widetilde V_{_{\rm p-p}}$), 
and hole-hole ($\widetilde V_{_{\rm h-h}}$); 
\begin{eqnarray}
   &&\hskip-1.0cm
  \left[i\hbar\;{\partial\over\partial t} - e_{ph}-e_{p'h'}\right]
  d_{pp',hh'} = 
   \nonumber \\ &&\hskip-1.0cm
    \bigl\langle \smf{p p'} \big|W\big| \smf{hp''}  \bigr\rangle\, c_{p''h'} +
    \bigl\langle \smf{p p'} \big|W\big| \smf{p''h'} \bigr\rangle\, c_{p''h}
   -\bigl\langle \smf{p h''} \big|W\big| \smf{hh'} \bigr\rangle\, c_{p'h''} -
    \bigl\langle \smf{h''p'} \big| W \big| \smf{hh'} \bigr\rangle\, c_{p h''}
   \nonumber \\
  &+&\!\!\biggl[
    \bigl\langle \smf{h''p'} \big|\Vphe\big| \smf{p''h'} \bigr\rangle_{\!_a\,}
    d_{pp'',hh''}
   -
    \bigl\langle \smf{h''p'} \big| \Vphe \big| \smf{h p''} \bigr\rangle_{\!_a\,}
    \,d_{pp'',h''h'} + \symph \; \biggr]
   \nonumber \\
   &+&\bigl\langle  \smf{pp'} \big| \widetilde V_{_{\rm p-p}} \big| \smf{p''p'''} \bigr\rangle
    \,d_{p''p''',hh'}
   +\bigl\langle \smf{h''h'''} \big|\widetilde  V_{_{\rm h-h}} \big| \smf{hh}' \bigr\rangle
    \,d_{pp',h''h'''}
\;.
\label{eq:2p2hweakc}
\end{eqnarray}
To simplify the calculation, we retain only those terms that survive
in the limit of a boson theory: This implies the neglect of exchange
effects and also of all contributions of the ``ladder'' type (which
are invoked through the Jastrow correlation operator $F$ and are
exactly accounted for in the Bose case). In the language of commonly
used Feynman diagrams this amounts to keeping the proper polarization
parts shown in Fig.~\ref{fig:FD_Pol}. Clearly, the graphs no.~3 and
5$-$7 invoke vertex corrections, which are beyond the CRPA formalism.

We also assume a {\it local\/} pair excitation operator,
{\it i.e.}\/
\begin{equation}
        d_{pp'hh'}\; a^\dagger_p a^\dagger_{p'}
        a^{\phantom\dagger}_{h'} a^{\phantom\dagger}_{h}
 \;\mbox{\small$\longrightarrow$}\;
 \sum_{qq'} d({\bf q},{\bf q}')
\left[\hat\rho_{\bf q}\hat\rho_{{\bf q}'}-\hat\rho_{{\bf q}+{\bf q}'}\right]\,,
\label{2.10}\end{equation}
where $\hat\rho_{\bf q}$ is the density operator,
and further simplify the $c$-$d$ coupling by replacing the
amplitudes $c_{ph}(t)$ by their Fermi-sea averages.
With these approximations the EOMs can be solved for $c_{ph}$ and $d_{ph,p'h'}$
which, in turn, yield the linear response wavefunction and thus the induced
density $\delta\rho({\bf r},t)$.

Introducing 
\begin{equation}
  \chi^{(\pm)}_0(q,\omega)=
  \pm{1\over N}\sum_h {n_h \, \overline{n}_{\bf{h+q}} \over
    \hbar\omega \mp e_{h+q,h}+ i0^+} \,,
\label{2.12}
\end{equation}
(where $n_k=\theta(k_{\rm F}\!-\!k)\equiv 1- \overline n_k$), the final
result for the density-density response function can be expressed as follows:
\begin{equation}
\chi^{\rm Pair}(q,\omega) = {\kappa(q,\omega)\over 1-
\kappa(q,\omega)\Vph{q}}
\label{2.14}\end{equation}
\begin{equation}\kappa(q,\omega) ={ \chi^{(+)}_0
\over 1-\chi^{(+)}_0 {\cal W}_+} +
{\chi^{(-)}_0 \over 1-\chi^{(-)}_0 {\cal W}_-}\,,
\label{2.15}\end{equation}
\begin{equation}
{\cal W}_{\pm} =
{\cal W}(q,\mp \omega) = {1\over2}\,
 \sum_{{\bf q}'} {\overline{\cal W}\!_3({\bf q}-{\bf q}',{\bf q}')\,
\overline {\cal W}\!_3({\bf q}-{\bf q}',{\bf q}') \over \mp\hbar\omega +
 \varepsilon({\bf q}-{\bf q}') +\varepsilon(q')}\;.
\label{2.16}\end{equation}
Here $\overline {\cal W}\!_3$
is a three-``phonon'' vertex, and $\varepsilon(q) = \hbar^2
 q^2/(2mS(q))$ is the Feynman spectrum. 

We conclude this section by noting that it is easily proved that
$\chi(q,\omega)$ obeys the $\omega^0$ and $\omega^1$ sum rules, {\it
independent}\/ of the specifics of ${\cal W}(q,\omega)$. This means
that $\Vph{q}$ is {\it uniquely\/} defined by the static structure
function through the first two energy weighted sumrules.

\section{RESULTS}
\label{sec:results}

\begin{figure}[h!]
  \includegraphics[width=.98\textwidth]{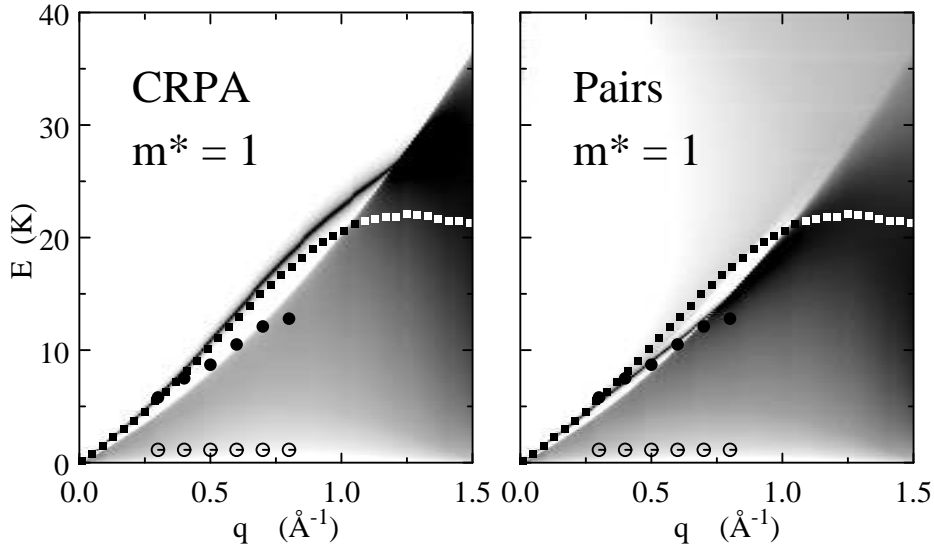}
  \caption{Saturation density $^3$He dynamic structure factor.
     Dark line: theoretical zero-sound mode.
     Filled [open] dots: experimental phonon [magnon]
     modes (Ref.~\protect\cite{GFvDG00}).
     Squares: RPA.}
  \label{fig:Sokw-both}
\end{figure}

In Fig.~\ref{fig:Sokw-both} we show the results for the dynamic
structure factor $S(k,\omega)$ of $^3$He.  Input to all our
calculations was the static $S(k)$ obtained in
Ref.~\onlinecite{polish}. The ``correlated RPA'' (= omitting pair
excitations) is shown on the left; as mentioned in the introduction
the collective mode is higher than found experimentally.  The right
picture gives the CBF result (= including pair fluctuations).
Obviously, the dispersion is significantly improved and a weak
pair-excitation background can be seen outside the patricle-hole
continuum.  As the calculated zero-sound mode is practically identical
with the experimental one, a modification of the effective mass is
unnecessary to obtain this result.

It appears interesting to investigate the effective mass of quasi%
--particle excitations in the system within this same formalism
and to compare the result with the phenomenological $m^*$ used in 
Ref.~\onlinecite{GFvDG00}. This can be done by studying the time
dependent wave function corresponding to a definite 
single particle-hole amplitude,
{\it i.e.}\/\ by omitting the sum over $ph$ (only) in the $c_{ph}-$%
contribution in Eq.~\ref{eq:FluctU}. Then one can proceed in exactly the
same way as above, leading to a consistent way of studying collective
excitations and quasi--particle properties. Work in this direction is
under investigation.

In summary, the introduction of dynamic multi-pair correlations, so
successful in $^4$He, also proves necessary for obtaining the correct
phonon dispersion in $^3$He.  A promising extension is the inclusion
of spin fluctuations, which holds the potential of understanding the
large effective mass in $^3$He.

As a matter of course, the theory can also be applied to other strongly
interacting Fermi fluids and to electrons.


\section*{ACKNOWLEDGMENTS}
This work was supported by the FWF project P18134-N08.
We thank H. Godfrin for
providing the data\cite{GodfrinMeschke}.


\begin{thebibliography}{8}

\bibitem{FeynmanBackflow}
R.~P. Feynman and M.~Cohen, \emph{Phys. Rev.}, \textbf{102}, 1189--1204
  (1956).

\bibitem{JacksonSumrules}
H.~W. Jackson, \emph{Phys. Rev. A}, \textbf{9}, 964--975 (1974).

\bibitem{VesaMikkou2}
V.~Apaja and M.~Saarela, \emph{Phys. Rev. B}, \textbf{57}, 5358 (1998).

\bibitem{GFvDG00}
H.~R.~Glyde, B.~F\aa{}k, N.~H.~van Dijk, H.~Godfrin, K.~Guckelsberger, and 
R.~Scherm, \emph{Phys. Rev. B}, \textbf{61}, 1421-1432 (2000).

\bibitem{IP84}
N.~Iwamoto and D.~Pines, \emph{Phys. Rev. B}, \textbf{29}, 3924-3935 (1984)

\bibitem{SiT81}
K.~S.~Singwi and M.~P.~Tosi, \emph{Solid State Phys.}, \textbf{36}, 177-266
(1981)

\bibitem{GodfrinMeschke}
H.~Godfrin, M.~Meschke, and H.~J. Lauter, (private communication).

\bibitem{FeenbergBook}
E.~Feenberg, \emph{Theory of Quantum Fluids}, New York: Academic Press,
1969.

\bibitem{polish}
E.~Krotscheck, \emph{J. Low Temp. Phys.}, \textbf{119}, 103 (2000).

\end{thebibliography}
\end{document}